\documentclass[epj]{svjour}
\usepackage{graphics}
\usepackage[isolatin]{inputenc}
\usepackage{dcolumn}
\begin{document}
\title{Precision mass measurements of radioactive nuclei at JYFLTRAP}
\author{S. Rahaman\thanks{\emph{E-mail address:} saidur.rahaman@phys.jyu.fi}, V.-V. Elomaa, T. Eronen, U. Hager, J. Hakala, A. Jokinen, A. Kankainen, 
I.D.~Moore, H. Penttilä, S. Rinta-Antila, J. Rissanen, A. Saastamoinen, T. Sonoda, C. Weber and J. Äystö 
%
}                     
%
%
\institute{Department of Physics, P.O. Box 35 (YFL), FIN-40014 University of Jyväskylä, Finland}
\date{Received: date / Revised version: date}
%
\abstract{The Penning trap mass spectrometer JYFLTRAP was used to measure the atomic masses of radioactive nuclei with an uncertainty better than 10 
keV. The atomic masses of the neutron-deficient nuclei around the \emph{N} = \emph{Z} line were measured to improve the understanding of the 
rp-process path and the SbSnTe cycle. Furthermore, the masses of the neutron-rich gallium (\emph{Z} = 31) to palladium (\emph{Z} = 46) nuclei have 
been measured. The physics impacts on the nuclear structure and the r-process paths are reviewed. A better understanding of the nuclear deformation is 
presented by studying the pairing energy around \emph{A} = 100.               
} 
\authorrunning{S. Rahaman, et al.}
\maketitle
\section{Introduction}
\label{intro}

 Penning traps are becoming an important device for experimental nuclear physics \cite{SI06,DL03}. The confinement of ions in a small volume by the 
combination of a homogeneous magnetic field and an electrostatic quadrupolar potential \cite{LB86} provides an ideal environment for performing mass 
measurements \cite{GB96} of nuclei far from the valley of stability. This offers a new opportunity for studying nuclear structure, stellar 
nucleosynthesis and weak interaction tests in nuclei \cite{KB06}. 

  This year a series of precise atomic mass excess values are reported in the vicinity of changes in the ground state nuclear structure 
\cite{UH06,UH06a,SR06}, astrophysics \cite{UH06a,SR06,AK06}, spectroscopy \cite{SRA06} and weak interaction studies \cite{TE06,TE06a}. References 
\cite{UH06,UH06a,SR06} provide accurate information concerning the mass surface and an onset of nuclear deformation in the region of 50 $\leq$ 
\emph{N} $\leq$ 82. These have been studied via the two-neutron separation and shell gap energies. This work substantially extends the previous 
knowledge and provides a better picture of the subshell closure and deformation effects in this region by studying the pairing energy.
          
  A further motivation for mass measurements at JYFL\-TRAP is to provide accurate mass values for a better understanding of the nucleosynthesis 
pathways \emph{i.e.} the rapid-proton capture process (rp-process) and the rapid-neutron capture process (r-process). In a first order approximation, 
the location of the rp-process and r-process paths are determined by the proton and neutron separation energies, respectively \cite{HS98,JC91}. Along 
the r-process path neutron separation energies are rather low, typically 2 to 3 MeV \cite{JC91}. The precise proton and neutron separation energies 
will help to pin down the location of the paths. Furthermore, in order to estimate the proton capture and neutron capture rates, the masses and 
excitation energies should be known to better than 10 keV.   

  The combination of the JYFLTRAP and the Ion Guide Isotope Separator On-Line (IGISOL) facility offers a unique opportunity to investigate the 
refractory elements and the medium-heavy nuclei. During the last two years the precision mass measurement program at JYFLTRAP has been applied to both 
neutron-rich and neutron-deficient nuclei for applications in different disciplines of physics. The limiting factors are the half-lives and the 
production rates of the radioactive nuclei. So far approximately 160 masses have been measured employing the JYFLTRAP setup. This article will give an 
overview of the JYFLTRAP mass data of interest for astrophysics and nuclear structure applications.
              
\section{The JYFLTRAP mass spectrometer}
\label{sec:2}


JYFLTRAP \cite{VK04,AJ06} is an ion trap experiment for cooling, bunching, isobaric purification and precision mass measurements of radioactive ions 
produced at the IGISOL facility \cite{JA01}. A schematic drawing of the JYFLTRAP setup together with the IGISOL facility is shown in Fig.~1. The 
radioactive nuclides are produced in proton-induced fission or heavy-ion fusion reactions by bombarding a thin target with a primary beam from the 
Jyväskylä K-130 cyclotron. Radioactive ions are extracted from the gas cell by helium gas flow and guided by the sextupole ion guide (SPIG) into a 
differential pumping stage where they are accelerated to 30 keV and mass-separated with a 55$^{0}$ dipole magnet. A mass resolving power ($M/\Delta 
M$) of up to 500 can be achieved. The production rates of the studied nuclei varied from 10,000 ions/s to 10 ions/s for the most exotic isotopes 
measured at the switchyard of IGISOL (see Fig.~1). The ions are transported to the radiofrequency quadrupole (RFQ) cooler \& buncher 
\cite{AN01,AJ02,JA03}. In this device the ions are cooled by collisions with helium buffer gas and are accumulated at the end of the RFQ structure.  
The ions are extracted at low energy as a short bunch with a time structure of $\sim$ 15 $\mu$s \cite{AN03} and are injected into the purification 
Penning trap, where the mass selective buffer gas cooling technique is applied for further cooling and isobaric cleaning \cite{GS91}. The mass 
resolving power of the purification trap is on the order of $10^{5}$. The purified and cooled ions are finally transported to the precision Penning 
trap where the cyclotron frequency ($\nu_{c}$) is measured by employing the time-of-flight technique \cite{MK95}. A typical time-of-flight resonance 
of $^{94}$Rb$^{+}$ radioactive ions is shown in Fig.~1. The cyclotron frequency is given by      
%
\begin{figure}
\resizebox{0.50\textwidth}{!}{%
\includegraphics{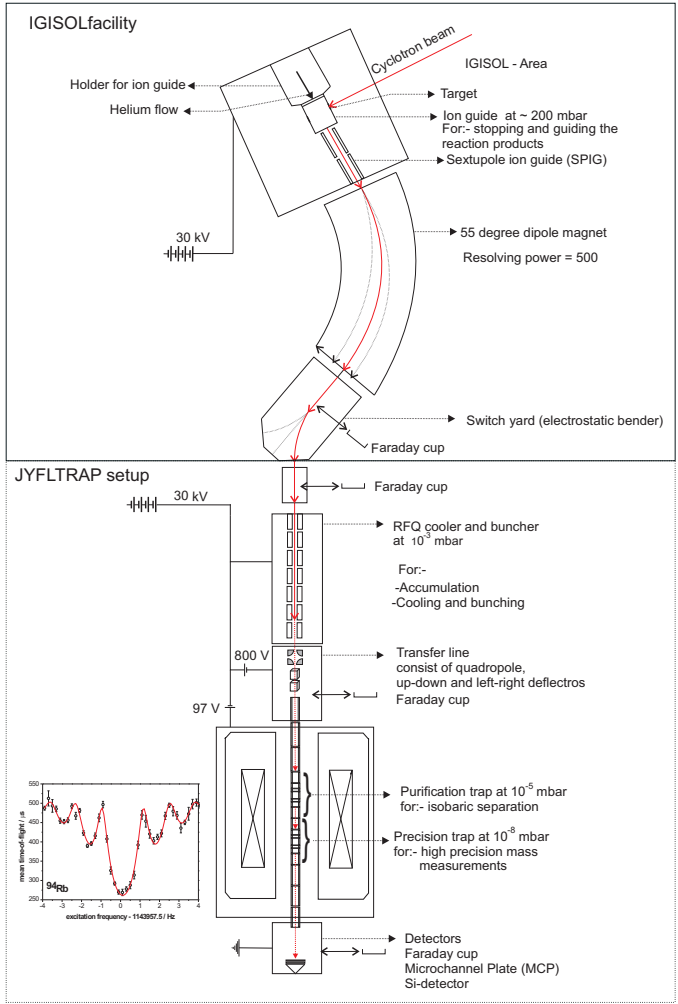}
}
\caption{A schematic drawing of the IGISOL facility and JYFLTRAP setup. The Penning traps are inside a single 7 Tesla superconducting solenoid.}
\label{fig:1}       
\end{figure}

\begin{equation}
\nu_{c} = \frac{1}{2\pi}\frac{q}{m}B,
\end{equation}
where \emph{B} is the magnetic field, \emph{m} is the mass and \emph{q} the charge state of the ion. To calibrate the magnetic field the cyclotron 
frequency ($\nu_{c}^{ref}$) of a precisely known reference mass ($m_{ref}$) is measured. The atomic mass of the ion of interest is then determined 
using the equation 
\begin{equation}
m = r \cdot(m_{ref} - m_{e}) + m_{e},
\end{equation}
where r = $\nu_{c}^{ref}$/$\nu_{c}$ is the frequency ratio and $m_{e}$ is the electron mass. 

In summary, a typical accuracy of 5 to 10 keV has been obtained in the reported experiments so far \cite{UH06,UH06a,SR06,AK06,SRA06}. An average 
transmission efficiency of the JYFLTRAP setup is 20\%. This has been measured for radioactive ions by comparing the number of detected beta decays at 
silicon detector 1 and 2, respectively (see Fig.~1).    
 

%
\section{Mass measurements at JYFLTRAP}
\label{sec:1}
\begin{table}
\begin{center}
\caption{List of the radioactive nuclei measured at JYFLTRAP to date.}
\label{tab:1}       
\begin{tabular}{llll}
\hline\noalign{\smallskip}
Element & Mass number & Reference \\
\noalign{\smallskip}\hline\noalign{\smallskip}
 Si & 26 & \cite{TE07}\\
 Al & 26 & \cite{TE06a,TE07}\\
 Sc& 42 & \cite{TE06a,TE07}\\
 Ti& 42 &\cite{TE07}\\
 V & 46 & \cite{TE06a}\\
 Cu& 62 & \cite{TE06}\\
 Zn & 62 & \cite{TE06}\\
 Ga & 62, 79-82 & \cite{TE06,JH07}\\
 Ge& 80-85 & \cite{JH07}\\
 As & 82-87 &\cite{JH07}\\
 Se & 85-89 & \cite{JH07}\\
 Br & 85-92 & \cite{SR06}\\
 Rb & 94-97 & \cite{SR06}\\
 Sr & 95-100 & \cite{UH06}\\
 Y& 79-83, 95-101 & \cite{AJ07,AK06}\\
 Zr& 85-86, 88, 98-105 &\cite{UH06,AK06}\\
 Nb & 85-88, 95-107 & \cite{SRA06,AK06,AJ07}\\
 Mo& 86, 88-89, 102-110 & \cite{UH06,VE07}\\
 Tc & 88-92, 106-112 & \cite{UH06a,VE07}\\
 Ru & 90-94, 106-114 & \cite{UH06a,VE07}\\
 Rh& 92-95, 108-118 & \cite{UH06a,VE07}\\
 Pd & 97-98, 112-120 &\cite{UH06a,VE07}\\
 Ag& 100 & \cite{VE07}\\
 Cd & 101-105 & \cite{VE07}\\
 In & 102,104&\cite{VE07}\\
 \noalign{\smallskip}\hline
\end{tabular}
\end{center}
\end{table}

Table~1 shows a list of the radioactive nuclei investigated at JYFLTRAP. To date 160 isotopes have been measured, contributing to the wide range of 
physics applications summarized in the introduction. Some of the works related to these data are presently under preparation to be published 
\cite{AJ07,JH07,VE07,TE07}. The precise \emph{Q} value measurements of the superallowed $\beta$ emitters can be found in refs. \cite{TE06,TE06a} and a 
summary can be found in \cite{CW06}.

\subsection{Masses for astrophysical applications}
In the early days of the JYFLTRAP experiments neutron-rich fission fragments were investigated. Recently, JYFL\-TRAP extended measurements to include 
fusion products towards the medium-heavy neutron-deficient region of the nuclear chart. This allows the investigation of nuclei that are important for 
the rp-process. 

The hot CNO cycle ignites the rp-process reaction and is the dominant reaction sequence for the nucleosynthesis and energy generation in hot stellar 
hydrogen burning. A long standing question of where the rp-process ends has been solved by the nucleosynthesis network calculations \cite{HS98} 
suggesting a termination at the SnSbTe cycle. A detailed view of the rp-process pathway is shown in Fig.~2 as predicted in ref. \cite{HS98}. Most of 
the masses along the rp-process path have not been determined experimentally. In particular, the masses of the tin, antimony and tellurium isotopes 
are very poorly known and prevent a reliable prediction of the endpoint of the path. The JYFLTRAP mass values \cite{AK06,VE07} in this region will 
therefore considerably improve the nucleosynthesis calculations.  

\begin{figure}
\resizebox{0.50\textwidth}{!}{%
\includegraphics{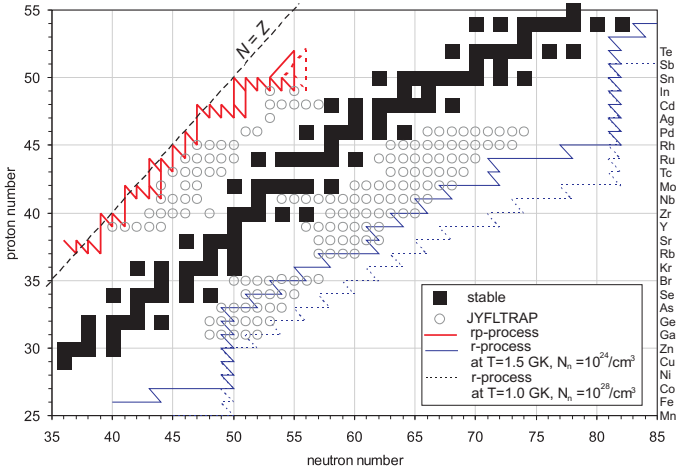}
}
\caption{A part of the nuclear chart is displayed for \emph{N} = 35 to 85 and \emph{Z} = 25 to 55. The black squares indicate the stable nuclei. The 
empty circles are the radioactive nuclei investigated at JYFLTRAP. A predicted rp-process pathway is marked by a thick line. Two possible r-process 
pathways are marked by the thin and dotted lines, respectively. }
\label{fig:2}       
\end{figure}

The rapid neutron-capture process (r-process) has a fundamental importance as it explains the origin of approximately half of the stable nuclei 
heavier than iron in nature. It occurs at the center of type II supernova at a very high temperature and neutron density. Two possible r-process 
pathways calculated according to the canonical model at two different conditions are shown in fig.~2 \cite{PA65}. Nuclear masses have the most 
decisive influence on the reaction flow of the r-process paths and they determine the neutron separation energies which characterize the paths. 
Therefore masses are among the most critical nuclear parameters in nucleosynthesis calculations \cite{JC91}. The r-process modeling is quite poorly 
known because of the high uncertainty in nuclear mass, often up to 1 MeV. The precise JYFLTRAP mass values close to the r-process path are reported in 
refs. \cite{UH06a,SR06,AJ07,JH07} and will substantially improve the r-process network calculations to minimize the ambiguity.         
          
\subsection{Masses for probing nuclear structure}

One recent highlight is the mass measurements of neutron-rich nuclei around \emph{A} = 100. The investigations are motivated due to theoretical 
\cite{KH04} and spectroscopic information on an onset of large deformation of the nuclei at \emph{N} $\geq$ 60. A systematic study of the two-neutron 
separation energies of a series of neutron-rich isotopes of rubidium, strontium, zirconium and molybdenum at JYFLTRAP reveals an onset of deformation 
and is reported in refs. \cite{UH06,SR06}. The extension of this knowledge of deformation to the higher and lower \emph{Z} regions is also reported in 
refs. \cite{UH06a,SR06}. 

 The two-neutron separation energy ($S_{2n}$) is shown as a function of neutron number in Fig.~3 for the even \emph{Z} nuclei. Generally, $S_{2n}$ 
decreases smoothly with neutron number and deformation effects appear as a discontinuity. The discontinuity in Fig.~3 confirms the maximum deformation 
in the nuclear shape for the zirconium isotope at \emph{N} = 59.       

\begin{figure}
\resizebox{0.50\textwidth}{!}{%
\includegraphics{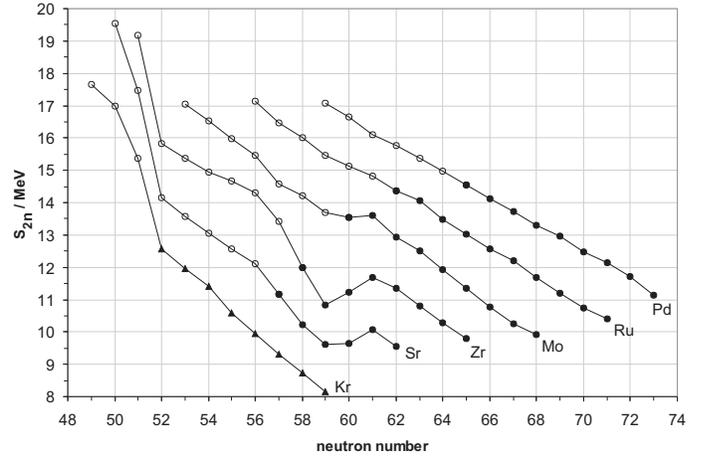}
}
\caption{Experimental two-neutron separation energy as a function of neutron number for even \emph{Z} nuclei. The filled and empty circles indicate 
the JYFLTRAP and the AME 2003 \cite{AME03} values, respectively. The values for krypton isotopes (filled triangles) are taken from ref. \cite{PD06}.}
\label{fig:3}       
\end{figure}

Another highlight is the shell gap quenching towards the lower \emph{Z} region for the \emph{N} = 50 magic shell closure. It has been predicted 
\cite{RC99,MS02} that the shell gap energy will be reduced towards \emph{Z} $\sim$ 28, but no experimental data has been available until now. A report 
\cite{SR06} from JYFLTRAP confirms the reduction of the shell gap energy towards $^{85}_{35}$Br. These measurements have been continued towards 
$^{83}_{31}$Ga at lower \emph{Z} \cite{JH07}.          
             
All articles cited above discuss the nuclear structure by studying the two-neutron separation energies and shell gap energies.
Further means of investigating the nuclear structure is via the neutron or proton pairing energy. To first order the pairing energy is calculated by 
using the formula 
\begin{equation}
P = \frac{12}{\sqrt{A}} ,
\end{equation}
in MeV, where \emph{A} is the mass number. The neutron pairing energy $\emph{P}_{3}(N)$ can also be calculated by using the three-point formula 
\cite{AME03} 
\begin{equation}
P_{3}(N) = \frac{(-1)^{N}}{4}[2 \cdot S_{n}(N) - S_{n}(N+1) - S_{n}(N-1)]
\end{equation}
and is shown in Fig.~4 for the strontium, zirconium, molybdenum and ruthenium isotopes. At \emph{N} = 56 an increase in the pairing energy is seen for 
strontium and zirconium due to the subshell closure. The plot also shows an enhancement of the pairing energy of about 50\% at \emph{N} = 59 compared 
to the neighbouring values. In the case of the subshell closure the pairing energy enhancement is seen due to a rapidly changing single-particle 
structure that results from the strong residual neutron-proton interaction between spin orbit partners. On the other hand, the enhancement of the 
pairing energy in the case of the deformed nuclei is not due to the single-particle structure but due to a weaker pairing. 
The addition of the JYFLTRAP precise mass values and hence the pairing energy values confirm the proposed theoretical prediction of relatively weaker 
pairing and hence higher pairing energy for the deformed nuclei \cite{JD95,RC91}.  

\begin{figure}
\resizebox{0.50\textwidth}{!}{%
\includegraphics{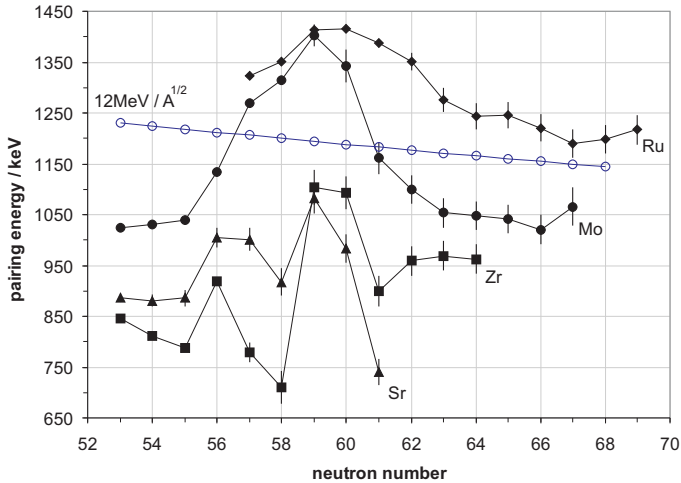}
}
\caption{Experimental pairing energy as a function of neutron number. The data are from the JYFLTRAP are complemented by the AME 2003 \cite{AME03}. 
The empty circles indicate the pairing energy calculated according to Eq.~3 for molybdenum isotopic chain.}
\label{fig:4}       
\end{figure}

\section{Conclusions and outlooks}
   
Precision mass measurements at JYFLTRAP and the applicability to very short-lived nuclei has been extended considerably. It is possible to reach a 
relative uncertainty better than $10^{-8}$ \cite{TE06a} and access to nuclei that are produced with a few tens of $\mu$b. The JYFLTRAP mass 
measurements cover a variety  of topics in nuclear physics, astrophysics and weak interaction studies with approximately 160 nuclei measured to date 
(see table~1). The carbon cluster ion source is currently under development and it will provide absolute mass determinations. Stabilization techniques 
will be implemented for both the pressure and temperature of the superconducting magnet to improve the stability of the magnetic field. In this way, 
along with the IGISOL developments \cite{IM05} the accessible range of nuclei will be further extended in the future.       

\begin{acknowledgement}
This work has been supported by the TRAPSPEC Joint Research Activity project under the EU 6th Framework program "Integrating Infrastructure Initiative 
- Transnational Access", Contract Number: 506065 (EURONS) and by the Academy of Finland under Finnish center of Excellence Program 2006-2011 (Nuclear 
and Accelerator Based Physics Program at JYFL) and project number 202256 and 111428.
\end{acknowledgement}
 
%

\end{document}